\documentclass{llncs}
\usepackage{amsmath}
\usepackage{graphicx}
\usepackage{subfigure}
\usepackage{ifthen}
\usepackage{enumitem}
\setlist{nolistsep} 

\def\MST{\mbox{MST}}
\def\OPT{\mbox{OPT}}
\def\Left{\mbox{left}}
\def\In#1{\mbox{in$(#1)$}} 
\def\Out#1{\mbox{out$(#1)$}} 
\def\Net#1{\mbox{net$(#1)$}} 
\def\xeP{x_{(e,P)}} 

\title{Light Spanners in Bounded Pathwidth Graphs}
\author{Michelangelo Grigni \and  Hao-Hsiang Hung}
\institute{Dept.\ of Math \&\ CS, Emory University,
\textsl{\{mic,hhung2\}@mathcs.emory.edu}}%

\begin{document}
\maketitle


\begin{abstract}

Given an edge-weighted graph $G$ and $\epsilon>0$, a
$(1+\epsilon)$-spanner is a spanning subgraph $G'$ whose shortest path
distances approximate those of $G$ within a factor of $1+\epsilon$.  
For $G$ from certain graph families (such as bounded genus graphs and apex
graphs), we know that \emph{light} spanners exist.  That is, we can
compute a $(1+\epsilon)$-spanner $G'$ with total edge weight at most a
constant times the weight of a minimum spanning tree.  This constant
may depend on $\epsilon$ and the graph family, but not on the
particular graph $G$ nor on the edge weighting.  The existence of
light spanners is essential in the design of approximation schemes for
the metric TSP (the traveling salesman problem) and similar
graph-metric problems.

In this paper we make some progress towards the conjecture that light
spanners exist for every minor-closed graph family: we show that light
spanners exist for graphs with bounded pathwidth, and they are computed
by a greedy algorithm.  We do this via the intermediate construction of
light \emph{monotone} spanning trees in such graphs.
\end{abstract}

\section{Introduction}

\subsection{Light Spanners}

Suppose $G$ is a connected undirected graph where each edge $e$ has
length (or weight) $w(e)\geq 0$.  Let $d_G (u, v)$ denote the length
of the shortest path between vertices $u$ and~$v$.  Suppose $G'$ is a
spanning subgraph of $G$, where each edge of $G'$ inherits its weight
from $G$; evidently $d_G(u,v) \leq d_{G'}(u,v)$.  Fix $\epsilon > 0$.
If $d_{G'}(u,v) \leq (1+\epsilon)\cdot d_G (u,v)$ (for all $u,v$),
then we say that $G'$ is a \emph{$(1+\epsilon)$-spanner} of $G$.  In
other words, the metric $d_{G'}$ closely approximates the
metric~$d_G$.

Let $w(G')$ denote the total edge weight of $G'$, and let $\MST(G)$
denote the minimum weight of a spanning tree in $G$.  We are
interested in conditions on $G$ that guarantee the existence of a
$(1+\epsilon)$-spanner $G'$ with bounded $w(G')/\MST(G)$.
Suppose $\mathcal{G}$ is a family of undirected graphs.  We
say \emph{$\mathcal{G}$ has light spanners} if the following
holds: for every $\epsilon>0$ there is a bound $f(\epsilon)$, so that
for any edge-weighted $G$ from $\mathcal{G}$, $G$ has a
$(1+\epsilon)$-spanner $G'$ with $w(G') \leq f(\epsilon)\cdot\MST(G)$.
Less formally, we say that $G'$ is a \emph{light spanner} for $G$.
Note $f(\epsilon)$ depends on $\epsilon$ and $\mathcal{G}$, but not on
$G$ or~$w$.

We know that if a graph family has unbounded clique minors, then it
does not have light spanners; just consider a clique with uniform edge
weights.  We conjecture the
converse~\cite{Grigni:2000:ATG:646253.686316}:
\begin{conjecture}\label{conj:main}
Any graph family with a forbidden minor has light spanners.
\end{conjecture}

Our pursuit of this conjecture is guided by the Robertson-Seymour
theory \cite{DBLP:journals/jct/RobertsonS03a}, which characterizes 
minor-closed graph families using four elements: bounded genus graphs,
apices, vortices, and repeated clique-sums.  We already know that if
$G$ has bounded genus or is an apex graph, then it has light
spanners~\cite
{Grigni:2000:ATG:646253.686316,Grigni:2002:LSA:545381.545492}.
In particular vortices are bounded pathwidth subgraphs, stitched inside
the faces of a bounded genus graph.

\subsection{Motivation}

Conjecture~\ref{conj:main} seems like a natural question, and its
proof would address the ``main difficulty'' discussed in the
concluding remarks of Demaine \emph{et
al.}~\cite{Demaine:2007:AAV:1283383.1283413}, in the general context
of approximation algorithms on weighted graphs.  As a specific
motivating problem, we review some results on the \emph{metric TSP},
the Traveling Salesman Problem with triangle inequality.  (For some
other problems, see
\cite{Berger05approximationschemes,DBLP:conf/icalp/BergerG07}.)

 We are given an edge-weighted graph $G$, and we seek a cyclic
order of its vertices with minimum total distance as measured by
$d_G$.  Equivalently, we want a minimum weight cyclic tour in $G$
visiting each vertex at least once.  Let $\OPT(G)$ denote the minimum
tour weight; it is well known that $\MST(G) \leq \OPT(G) \leq
2\cdot\MST(G)$.  We seek an approximation scheme: an algorithm which
takes as inputs the weighted graph $G$ and $\epsilon>0$, and which
outputs a tour with weight at most $(1+\epsilon)\cdot\OPT(G)$.

The problem is MAX SNP-hard~\cite{PY-tsp12-93}, so we consider
approximation schemes where the input graph $G$ is restricted to some
graph family $\mathcal{G}$ (e.g., planar graphs).  We would like a
PTAS (an approximation scheme running in time $O(n^{g(\epsilon)})$,
for some function $g$), or better yet an EPTAS (an approximation
scheme running in time $O(g(\epsilon)\cdot n^c)$, where the constant
$c$ is independent of $\epsilon$).

Suppose $\mathcal{G}$ is a graph family, and that for any $G \in
\mathcal{G}$ we can compute a $(1+\epsilon)$-spanner $G'$ with $w(G')
\leq f(\epsilon)\cdot \MST(G)$.  Then we may attempt to design a PTAS
(or an EPTAS) for the metric TSP on $\mathcal{G}$, as follows:
\begin{enumerate}

\item On input $G$ and $\epsilon$, first compute $G'$, a
  $(1+\epsilon/2)$-spanner of $G$, with weight at most
  $f(\epsilon/2)\cdot \MST(G)$.

\item Choose $\delta = (\epsilon/2)/f(\epsilon/2)$.
Apply some algorithm finding a tour in $G'$ with cost at most
$\OPT(G') + \delta \cdot w(G')$.

\item Return the tour, with cost at most $(1 + \epsilon/2)\cdot\OPT(G)
+ \delta \cdot (f(\epsilon/2)\cdot\MST(G)) \leq
(1+\epsilon)\cdot\OPT(G)$.  (For other metric optimization problems,
it may be less trivial to lift a solution from $G'$ back to $G$.)

\end{enumerate}
Step 2 looks like the original problem, except now we allow an error term
proportional to $w(G')$ instead of $\OPT(G')$.   This
approach has already succeeded for planar graphs~\cite
{DBLP:conf/soda/AroraGKKW98,KleinTSP2005} and bounded genus
graphs~\cite
{Demaine:2007:AAV:1283383.1283413,Grigni:2000:ATG:646253.686316}.

A recent result of Demaine \textit{et
al.}~\cite[Thm.~2]{ContractionMinorFree_STOC2011} implies a PTAS for
metric TSP when $\mathcal{G}$ is \emph{any} graph class with a fixed
forbidden minor.  Since we do not know that $\mathcal{G}$ has light
spanners, for step~1 they substitute a looser result~\cite
{Grigni:2002:LSA:545381.545492}, finding a $(1+\epsilon)$-spanner $G'$
with weight $O((\log n )/\epsilon)\cdot\MST(G)$ (the hidden constant
depending on $\mathcal{G}$).  In step~2 their algorithm runs in time
$2^{O(1/\delta + \log n)}$. Their $1/\delta$ is
$O(w(G')/(\MST(G)\cdot\epsilon))=O((\log n)/\epsilon^2)$, so their
running time is $n^{O(1/\epsilon^2)}$.  If we could compute light
spanners for $\mathcal{G}$, then $\delta$ would improve to something
independent of $n$, and this would yield an EPTAS for metric TSP on
$\mathcal{G}$. (Or alternatively, it would yield an approximation
scheme allowing $\epsilon$ to slowly approach zero, as long as
$1/\delta$ stays $O(\log n)$.)

\subsection{Our Work}

In this paper we make some progress towards
Conjecture~\ref{conj:main}: we show that light spanners exist for
bounded pathwidth graphs.
\begin{theorem}\label{thm:main}
Bounded pathwidth graphs have light spanners, computable by a greedy
algorithm.
\end{theorem}
We prove this in Section~\ref{sec:mainarg}.
This result is not algorithmically interesting by itself, since metric
TSP (and many other problems) is exactly solvable in polynomial time
when $G$ has bounded pathwidth, or even bounded treewidth.  Rather, we
regard this as progress towards the conjecture, and towards an EPTAS
for metric TSP (and similar problems) on graphs with forbidden minors.
See Section~\ref{sec:further} for some further remarks.

\section{Preliminaries}

\subsection{Charging Schemes}

In order to exhibit light spanners in a weight-independent way, we use
\emph{charging schemes}~\cite{Grigni:2002:LSA:545381.545492}.  (We use
the notion called ``0-schemes'' in
\cite{Grigni:2002:LSA:545381.545492}, not the more general
``$\epsilon$-schemes'' required for apex graphs.)  Suppose each edge
of graph $G$ can hold some quantity of \emph{charge}, initially zero.
A \emph{detour} is an edge $e\in E$ and a path $P$ such that $e+P$ is
a simple cycle in $G$.  For each detour $(e,P)$ we introduce a
variable $\xeP\geq0$.  Each $\xeP$ describes a
\emph{charging move}: it subtracts $\xeP$ units of charge from
edge $e$, and adds $\xeP$ units of charge to each edge of $P$.
When $\xeP>0$, we say ``$e$ charges $P$''.

Given graph $G$, a spanning tree $T$, and a number $v$, a
\emph{charging scheme from $G$ to $T$ of value $v$} is an assignment
of nonnegative values to the $\xeP$ variables (i.e., a fractional
sum of detours) meeting the three conditions listed below.  Here 
$\Out{e}$ denotes the total charge subtracted from edge $e$,
$\In{e}$ denotes the total charge added to $e$ (as part of various detour
paths),
and $\Net{e} = \In{e} - \Out{e}$ is the total charge on $e$ after all
the moves are done:
\[
\begin{array}[t]{l@{~~~}rcl@{~~}l}
(1) & \Out{e} & \geq & 1
& \mbox{for all } e \in G-T, \\
(2) &     \Net{e} & \leq & 0 
& \mbox{for all } e \in G-T, \\
(3) &     \Net{e} & \leq & v
& \mbox{for all } e \in T.
\end{array}
\]
Note ``$e \in G-T$'' means $e$ is an edge of $G$ but not $T$.
As we'll see in Theorem~\ref{thm:greedy}, charging schemes imply light
spanners.
\begin{definition} 
An \emph{acyclic scheme} is a charging scheme with two additional
conditions:
\begin{description}
\item[\textnormal{(4)}] If edge $e$ charges some path, then $e\in
G-T$.
\item[\textnormal{(5)}] There is an ordering of the edges such that whenever
edge $e_1$ charges a path containing edge~$e_2$, $e_1$ precedes~$e_2$.
\end{description}
\end{definition}
For example, planar graphs have integral acyclic
schemes of value $v=2$~\cite{Althofer:1993:SSW:156252.156258}.
\begin{definition}
Suppose we have detours $(e_1,P_1)$ and $(e_2,P_2)$, with $e_2\in P_1$
and $e_1\not\in P_2$.  Their \emph{shortcut} is the detour $(e_1,P')$,
where $P'$ is the path derived from $P_1$ by replacing $e_2$ with
$P_2$, and then reducing that walk to a simple path.
\end{definition}

\begin{lemma}\label{lem:shortcut} 
Suppose we have an acyclic scheme of value $v$ from $G$ to $T$, and
an edge $e$ in $G-T$.  Then there is an acyclic scheme of value $v$
from $G-e$ to $T$.
\end{lemma}

\begin{proof}
Let $e_2=e$.  While $\In{e_2}$ is positive, we find some $e_1$ charging
a path $P_1$ containing $e_2$.  Since $\Net{e_2}
\leq 0$, $e_2$ also charges some path $P_2$.  $P_2$
cannot contain $e_1$, since the scheme is acyclic.  Let
$\alpha=\min(x_{(e_1,P_1)}, x_{(e_2,P_2)})$.  Now reduce both
$x_{(e_1,P_1)}$ and $x_{(e_2,P_2)}$ by $\alpha$, and increase
$x_{(e_1,P')}$ (their shortcut) by $\alpha$.  After this change all the
conditions are still satisfied, except possibly for condition (1) at $e_2$.
Repeat until $\In{e_2}$ reaches zero.  Finally remove $e_2$ and any
remaining charges out of~$e_2$.  \qed
\end{proof}

\begin{theorem}
\label{thm:greedy}
Suppose $G$ is a graph with spanning tree $T$, and we have an acyclic
scheme from $G$ to $T$ of value $v$.  Then for any $\epsilon > 0$, and
for any non-negative edge-weighting $w$ on $G$, a simple greedy
algorithm finds a $(1+\epsilon)$-spanner $G'$ in $G$ containing $T$,
with total weight $w(G')\leq(1+v/\epsilon)\cdot w(T)$.
\end{theorem}
We use the following greedy algorithm of Alth\"{o}fer 
\emph{et al.}~\cite{Althofer:1993:SSW:156252.156258},
modified to force the edges of $T$ into~$G'$:
\begin{quote}
\begin{tabbing}
xxx \= xxx \= xxx \= \kill
Spanner($G$, $T$, $1+\epsilon$): \\
\> $G' = T$ \\
\> for each edge $e\in G-T$, in non-decreasing $w(e)$ order \\
\>\> if $(1+\epsilon)\cdot w(e) < d_{G'}(e)$ then \\
\>\>\> add edge $e$ to $G'$\\
\>return $G'$
\end{tabbing}
\end{quote}
The proof of Theorem~\ref{thm:greedy} is a variant of previous 
arguments by LP duality~\cite
{Grigni:2000:ATG:646253.686316,Grigni:2002:LSA:545381.545492},
for completeness we sketch it here.
\begin{proof}
Since $G'$ is computed by the greedy algorithm, it is clearly a
$(1+\epsilon)$-spanner of $G$ containing $T$; the issue is to bound
its weight $w(G')$.  
By Lemma~\ref{lem:shortcut} we have an acyclic scheme from $G'$ to
$T$ of value~$v$.  

Consider a detour $(e,P)$ in $G'$ with $e\not\in
T$. We claim $(1+\epsilon)\cdot w(e) < w(P)$ (to see this, compare $e$
with the last edge inserted by the algorithm on the cycle $e+P$). 
Multiply through by $\xeP$ and we have this:
\begin{eqnarray*}
\xeP \cdot \epsilon \cdot w(e) & \leq & \xeP \cdot (w(P)-w(e))
\end{eqnarray*}
When $e\in T$ this is still valid, since $\xeP=0$.  Now sum over all
detours $(e,P)$:
\begin{eqnarray*}
\sum_{(e,P)} \xeP \cdot \epsilon \cdot w(e)
& \leq &
\sum_{(e,P)} \xeP \cdot (w(P)-w(e)) \\
\epsilon \cdot \sum_{e\in G'} w(e)\cdot \Out{e}
& \leq &
\sum_{e\in G'} w(e) \cdot \Net{e} \\
\epsilon \cdot w(G'-T) & \leq & v \cdot w(T)
\end{eqnarray*}
So $w(G') = w(T)+ w(G'-T) \leq (1+v/\epsilon)\cdot w(T)$.
\qed
\end{proof}

\subsection{Bounded Pathwidth and Monotone Trees}
\label{def:bpmt}
Suppose $G=(V,E)$ is a graph, $P$ is a path (disjoint from $G$), and
$\mathcal{B}={(B_i)}_{i\in P}$ is a collection of subsets of $V$ (bags) indexed
by vertices $i$ in $P$.  We call the pair $(P, \mathcal{B})$ a
\emph{path decomposition} of $G$ if the following conditions hold: (1)
${\bigcup}_{i\in P} B_i = V$; (2) for every edge $\{u, v\}\in E$, there
is at least one bag $B_i$ with $\{u,v\}\subseteq B_i$; (3) for
every $v\in V$, $\{i:\; v\in B_i\}$ is connected
(an interval) in $P$.  The \emph{pathwidth} of the decomposition is
the maximum bag size minus one, and the pathwidth of $G$ is the
minimum pathwidth of any path decomposition of $G$.

Given $(P, \mathcal{B})$, we may lay out $P$ on the line, and regard
$G$ as a subgraph of an interval graph.  That is, for each vertex $v$
we have a line interval $I_v$ (corresponding to an interval in $P$),
and we have $I_u \bigcap I_v\neq\emptyset$ whenever $\{u, v\}\in E$,
and at most $k+1$ intervals overlap at any point of the line.  For
convenience we may eliminate ties via small perturbation, so that all the interval endpoints
are distinct.  In particular, let $\Left(v)$ denote the leftmost point
of $I_v$.   


Suppose $T$ is a rooted tree in $G$.  We say $T$ is a \emph{monotone tree} if
for every vertex $v$ in $T$ with parent $p$, we have $\Left(p) <
\Left(v)$.  When $T$ is a path rooted at an endpoint, we say it is a
\emph{monotone path}.  In particular if $T$ is a monotone spanning
tree in $G$, then from any vertex $v$, we can find a monotone path in
$T$ from $v$ to the root of $T$ (the vertex with the leftmost
interval).  For this process, it is convenient to imagine that edges
connect intervals at their leftmost intersection point.

\section{Main Argument}
\label{sec:mainarg}

We are given $\epsilon>0$, a connected edge-weighted graph $G$ with
$n$ vertices, and an interval representation $\{I_v\}$ of $G$ with
pathwidth $k$.  We want to find a $(1+\epsilon)$-spanner $G'$ in $G$
of low weight.  First we apply some reductions to simplify $G$:

\emph{Nice Decomposition}. We may assume that each pair of consecutive bags (as vertex
sets) differ by only one vertex.  This can be enforced by an argument
similar to the construction of \emph{nice} tree-decompositions
\cite{DBLP:books/sp/Kloks94}: if two consecutive bags differ on $m\geq
2$ vertices, we introduce $m-1$ intermediate bags, in such a way that
each pair differs on only one vertex, and we do not increase the
maximum bag size. This does not modify $G$ at all.

\begin{figure}[ht]
\begin{center}
{\includegraphics[width=6cm]{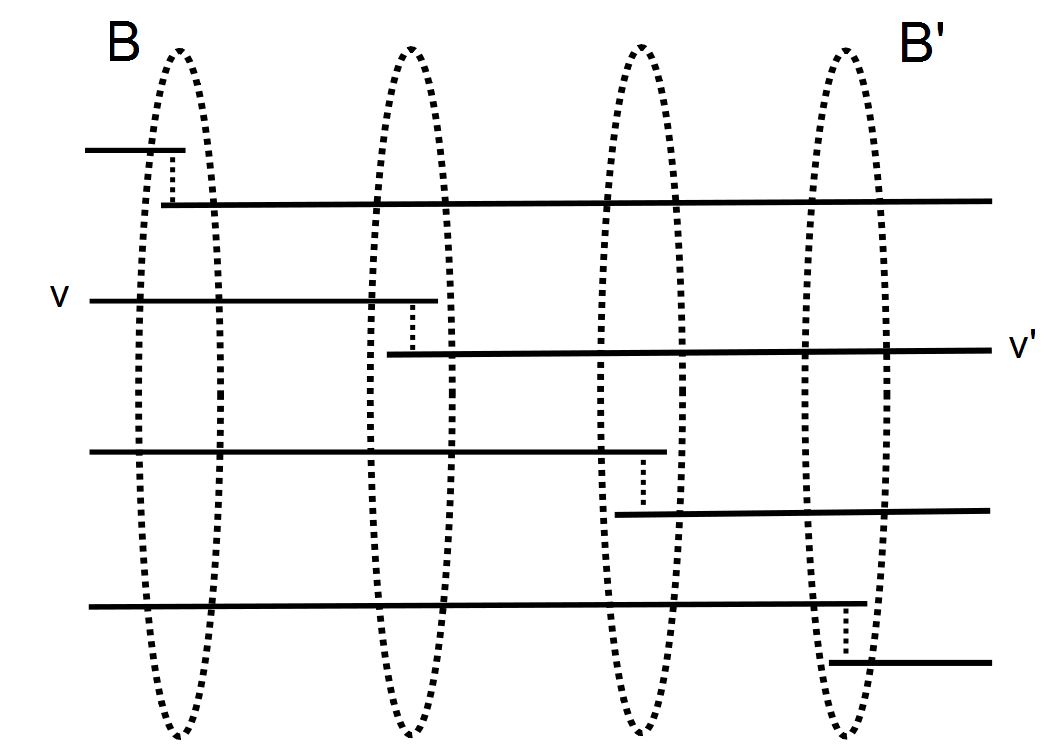}}
\caption{Each vertex $v$ in bag $B$ is replaced by $v'$ in bag $B'$.}
\label{fig:nicepath}
\end{center}
\end{figure}

\emph{Bounded Degree Assumption}. We may assume each vertex appears in $O(k)$ bags, and so the
maxdegree of $G$ is $O(k)$.  To enforce this, we copy the bags of $G$
from left to right.  After each group of $k$ original bags, ending
with a bag $B$, we insert $|B|$ ``replacer'' bags, each of which
replaces one vertex $v\in B$ with a copy $v'$, connected to $v$
by an edge of length zero. This ends with a bag $B'$, where every
vertex $v \in B$ has been replaced by a copy $v'\in B'$.
See Figure~\ref{fig:nicepath}.
We continue in this way (using the copies in place of the originals)
across the entire path decomposition.  If we aren't careful the
pathwidth may increase by one, but this does not matter for our
asymptotic results.  The original graph is obtained by contracting a
set $S$ of weight-zero edges in the modified graph.  So given a
spanner $G'$ in this modified graph, we may contract $S$ in $G'\cup S$
to recover a spanner (of no greater weight) in the original.

\emph{Completion Assumption}. We may assume that $G$ is \emph{completed}; that is, it contains
all edges allowed by its overlapping intervals.  In other words: we
have a clique in each bag, $G$ is an interval graph.  For each absent
edge $e=\{u,v\}$, we simply add it with weight $w(e)$ equal to the
shortest path length $d_G(u,v)$.  This does not change $d_G$ at all.
Given a spanner $G'$ in the completed graph, we recover a spanner in
the original graph by replacing each completion edge by the
corresponding shortest path.

\begin{proof}[of Theorem~\ref{thm:main}]
We assume all the above reductions have been applied: the input graph $G$
is a connected edge-weighted interval graph of width $k$, each bag in its path 
decomposition introduces at most one vertex, and each vertex of $G$ 
has degree $O(k)$.

By Lemma~\ref{lem:monoT} (below), we compute a monotone spanning tree
$T$ with $w(T)=O(k^2)\cdot \MST(G)$.
By Lemma~\ref{lem:T2scheme} (below), we exhibit an acyclic charging
scheme from $G$ to $T$ of value $v=O(k)$.
Finally we apply the greedy algorithm, which computes
a ($1+\epsilon$)-spanner $G'$.
By Theorem~\ref{thm:greedy}, $w(G') \leq (1+v/\epsilon)\cdot w(T) =
O(k^3/\epsilon)\cdot \MST(G)$.
\qed
\end{proof}

\begin{lemma}\label{lem:monoT} Given $G$ as above, it contains a monotone
 spanning tree $T$ with $w(T)\leq O(k^2)\cdot \MST(G)$. 
\end{lemma}
\begin{proof}
Choose a minimum spanning tree $T^*$, so $w(T^*)=\MST(G)$.
Let $I_l$ and $I_r$ be the leftmost and rightmost intervals.
Let $P_1$ be a shortest path from $I_l$ to $I_r$; since $G$ is completed,
we may assume $P_1$ is monotone, as in Figure~\ref{fig:tree1}.
Note $w(P_1)\leq w(T^*)$.

Consider the components ${T_1}^*, {T_2}^*, ..., {T_m}^*$ of $T^* - V(P_1)$.
Let $e_i$ be an edge connecting the leftmost point of $T_i^\ast$ to a
vertex of $P_1$ (it exists by completion).
For each ${T_i}^*$, we recursively compute a monotone spanning tree $T_i$
of $G[V({T_i}^*)]$. Finally, $T = P_1 \cup \bigcup_i (T_i\cup e_i)$.

It is clear that $T$ is monotone, but we must account for the total
weight of $w(T)$.  For each component ${T_i}^*$, let $f_i$ be an edge
of $T^*$ connecting ${T_i}^*$ to $P_1$ (there must be at least one).
By triangle inequality, we see $w(e_i)$ is at most $w({T_i}^*) +
w(f_i) + w(P_{1,i})$, where $P_{1,i}$ is a subpath of $P_1$ from
the endpoint of $e_i$ to the endpoint of $f_i$.  Note the $f_i$'s and
$T^*_i$'s are disjoint parts of $T^*$, but the subpaths may overlap inside
$P_1$.

\begin{figure}[ht]
\begin{center}
{\includegraphics[width=7cm]{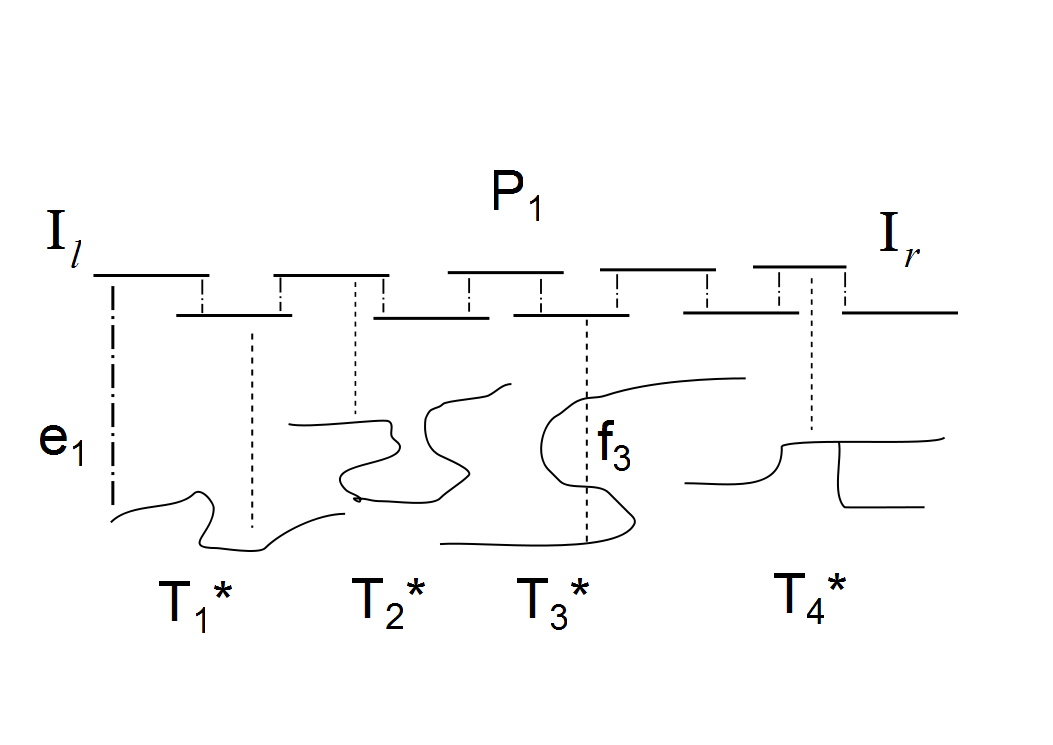}}
\caption{$P_1$ and the ${T_i}^*$ subtrees. Each 
$f_i$ in $T^*$ is replaced by an $e_i$ in $T$.}
\label{fig:tree1}
\end{center}
\end{figure}

An edge $e\in P_1$ appears in at most $k-1$ of the
$P_{1,i}$ subpaths, since each subpath witnesses another vertex (from
$T^*_i$) that must appear in the bag with~$e$.  So $\sum_i
w(e_i)\leq\sum_i [w(f_i)+w({T_i}^*)+w(P_{1,i})]\leq
w(T^*)+(k-1)w(P_1)\leq k\cdot w(T^*)$.  Since $w(T) \leq O(k\cdot
w(T^*)) + \sum_i w(T_i)$ and $\sum_i w({T_i}^*) \leq w(T^*)$, a simple
depth-$k$ recursion finishes our bound.  \qed
\end{proof}

\noindent\emph{Remark:} we do not have to compute $T$ as in
Lemma~\ref{lem:monoT}; it suffices to use any light enough monotone
spanning tree.  A natural choice is to let $T$ be the \emph{lightest}
monotone spanning tree, which we compute as follows.  Start with just
the root (in the leftmost bag), and grow the tree in a left-to-right
scan of the bags: each time a bag $B$ introduces a new vertex $v$, add
an edge connecting $v$ to its nearest neighbor in $B$ (which is
already in~$T$).

\begin{figure}[ht]
\begin{center}
\mbox{
\subfigure[a monotone tree $T$]{\includegraphics[width=5cm]{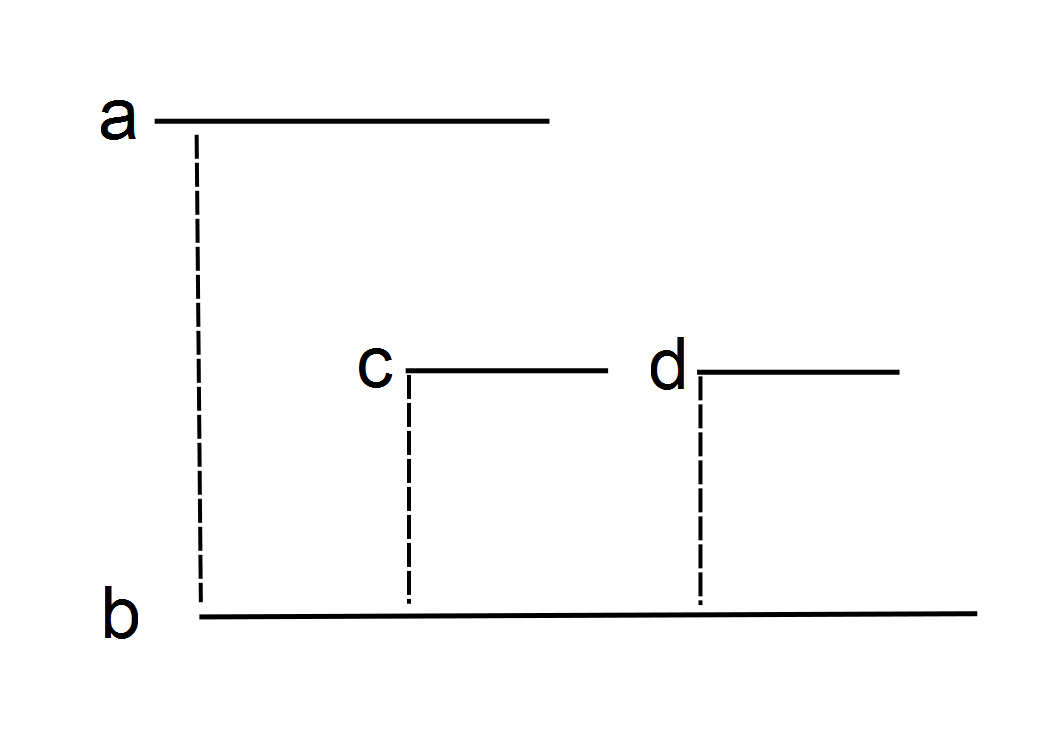}}
\subfigure[$T^{(2)}$ produced from $T$]{\includegraphics[width=5cm]{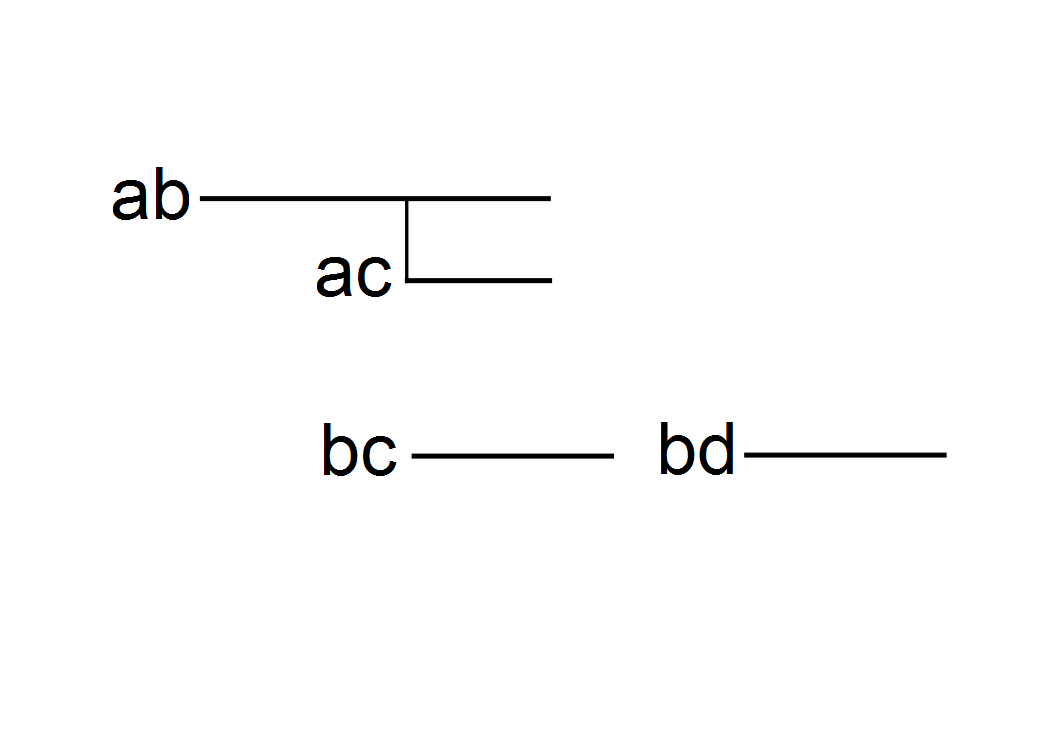}}
}
\caption{Horizontal lines are intervals, dashed verticals are edges of $T$.}
\label{fig:pages}
\end{center}
\end{figure}

In the completed $G$, a \emph{triangle move} is a charging move where
a non-tree edge $e$ charges a path $P$ of length two, where at most
one edge of $P$ is not in $T$.  We now define $T^{(2)}$, a graph whose
edges represent triangle moves.  Each vertex $jk$ of $T^{(2)}$
corresponds to an edge $\{j,k\}$ in $G$.  We also represent the vertex
$jk$ by the interval $I_{jk} = I_j \cap I_k$.  To define the edges of
$T^{(2)}$, we first define a \emph{parent} for each vertex $jk$.  If
$\{j,k\}$ is an edge of $T$, then $jk$ has no parent.  Otherwise,
suppose $\Left(j)<\Left(k)$ (else swap them), and let $i$ be the
parent of $k$ in $T$; $i$ must exist since $k$ is not the root.  Note
$\{i,j,k\}$ is a triangle in $G$.  Now we say the parent of $jk$ is
$ij$, and we add the edge $\{ij,jk\}$ in $T^{(2)}$.  Note
$\Left(ij)<\Left(jk)$, so these parent links are acyclic. Thus
$T^{(2)}$ is a forest, with each component rooted at a vertex
corresponding to an edge of $T$.  Figure~\ref{fig:pages} illustrates a
simple monotone tree $T$ and its forest $T^{(2)}$.

\begin{lemma}\label{lem:T2scheme} Given $G$ as above with a monotone
spanning tree $T$, there is an acyclic 
charging scheme from $G$ to $T$ of value $O(k)$.
\end{lemma}

\begin{figure}[ht]
\begin{center}
\mbox{%
\subfigure[some edges of T (solid) and G-T (dashed)]{\includegraphics[width=7cm]{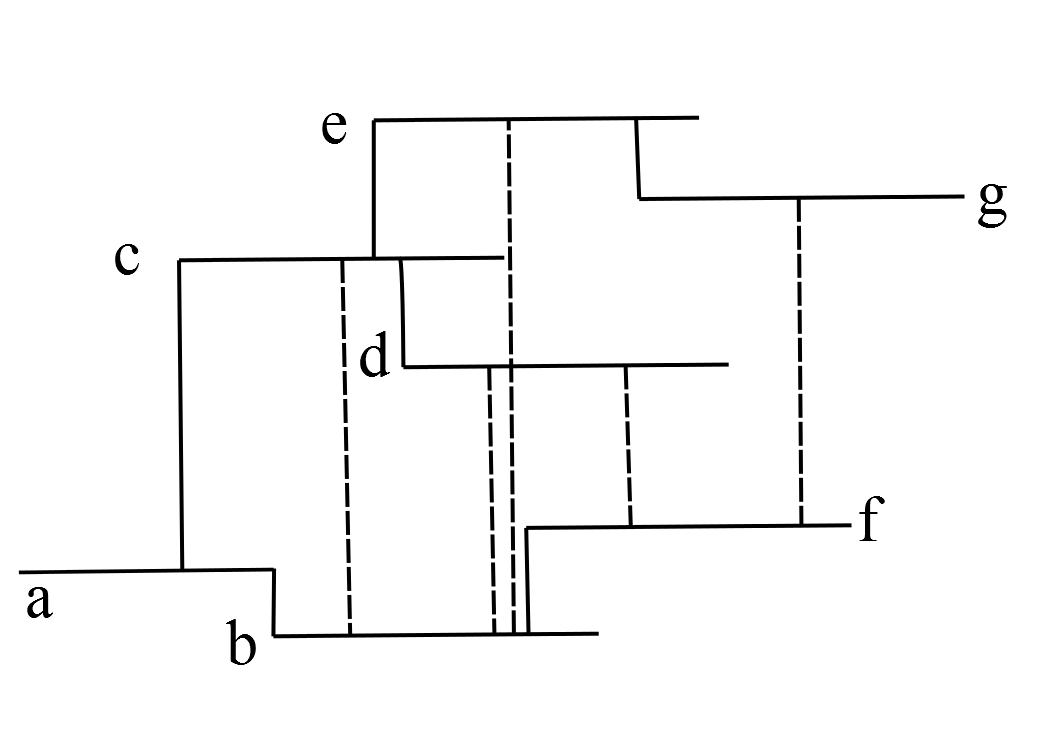}}
\subfigure[a shortcut tour in $T^{(2)}$, ending at ac]{\includegraphics[width=7cm]{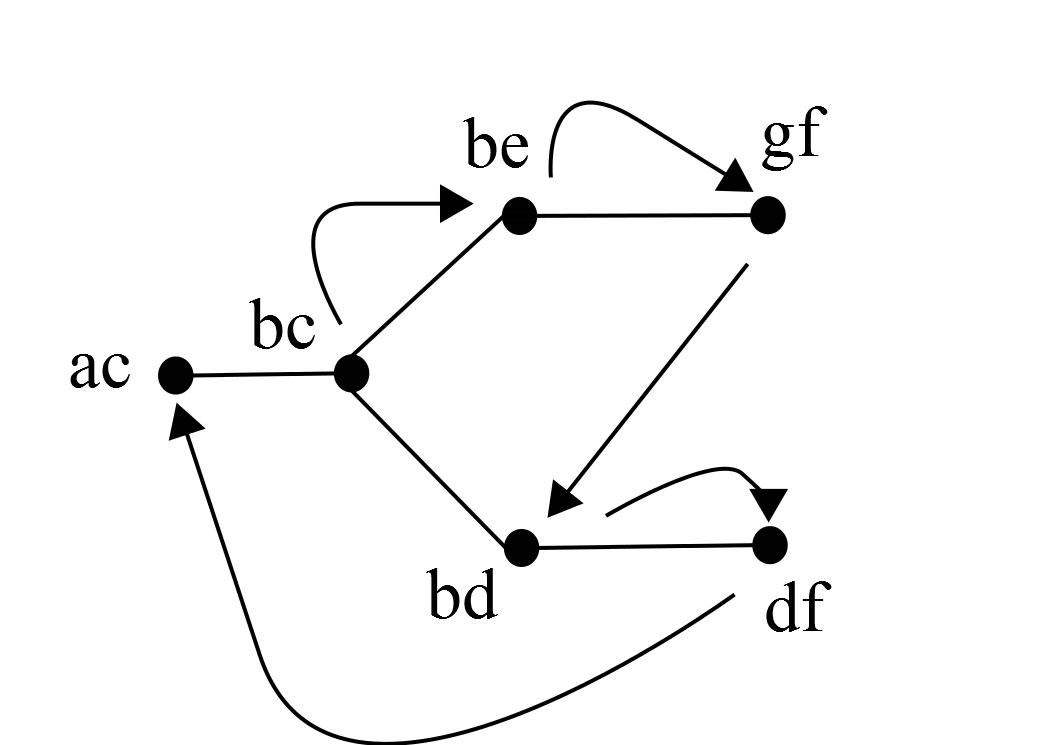}}}
\caption{Edges of $G$ (left) are vertices of $T^{(2)}$ (right)}
\label{fig:trees}
\end{center}
\end{figure}

\begin{proof}
Recall $T^{(2)}$ is a forest.  Fix a component $C$ of $T^{(2)}$; it is
a tree, rooted at a vertex $r$ corresponding to an edge of $T$, and
that is the only such vertex in $C$.  Consider a directed Euler tour
of $C$, traversing each edge twice.  Delete each tour edge out of $r$,
so we get a list of directed paths, each of the form
\[
e_1 \rightarrow e_2 \rightarrow \cdots e_m \rightarrow r
\]
where each vertex $e_i$ corresponds to some edge of $G-T$.  Since $C$
is a tree, these paths are vertex disjoint (except at $r$).  However,
a vertex may appear more than once on the same path; call an
appearance $e_i$ a \emph{repeat} if the same vertex appeared earlier
on the path.  Let $\mathcal{P}$ be the collection of all these paths,
from all components of $T^{(2)}$.

We now propose a charging scheme (which fails to be acyclic).  Recall
how we constructed edges in $T^{(2)}$: we connect each vertex $jk$
(corresponding to an edge of $G-T$) to its parent $ij$.  If a path in
$\mathcal{P}$ traverses this edge in the direction $jk \rightarrow
ij$, we add the triangle move where edge $\{j,k\}$ charges one unit
to path $j-i-k$.  If a path traverses this edge in the other
direction $ij \rightarrow jk$ (so $ij$ is not a tree edge), we
add the triangle move where edge $\{i,j\}$ charges one unit to path $i-k-j$.
In either direction, the tree edge $\{i,k\}$ is charged.

For an edge $e \in G-T$, the corresponding vertex appears at least
once on a path, and it has at least as many out-edges as in-edges, so
our proposed scheme satisfies conditions (1) and (2).  For an edge $e \in
T$, we must bound the number of times it is charged.  Since $G$ has
maxdegree $O(k)$, $e$ appears in $O(k)$ distinct triangles, and it is
charged at most twice per triangle (this includes the charges it
receives in its role as $r$).  So if we choose $v=O(k)$, condition (3) is satisfied.
Also there are no charges out of tree edges, so condition (4) is also satisfied.

However, this charging scheme does not satisfy condition (5); if a vertex
(corresponding to an edge $e \in G-T$) has a repeat appearance on its
path, then there is no consistent way to order the edges.  To fix
this, we eliminate all ``repeat'' appearances using shortcuts.
That is, whenever we have a sequence $e_1 \rightarrow e_2 
\rightarrow e_3$ where $e_2$ is a repeat, we shortcut out $e_2$.
Note such shortcuts can be combined.  For example if we have a sequence
$e_1 \rightarrow e_2 \rightarrow e_3 \rightarrow e_4 \rightarrow e_5$,
corresponding to four triangle moves, it is possible to shortcut out
$e_2,e_3,e_4$ (in any order), and the result is a single charge from
$e_1$ to a path containing $e_5$ (the rest of the charged path is all
tree edges).  After eliminating all repeats by shortcuts, we get the
desired acyclic scheme.
\qed
\end{proof}

\section{Conclusion and Further Work}
\label{sec:further}

Regarding our main result, it is not clear whether we really need
to force the edges of a monotone $T$ in the greedy spanner
computation.  Also, we might hope to reduce the $O(k^3)$ factor to
something smaller.  

The next obvious target is bounded treewidth graphs, a prerequisite
for handling clique sums as in the Robertson-Seymour characterization.

There are several obvious directions to try extending the current approach to
further minor-closed graph families.  First, as extensions of
Theorem~\ref{thm:main}, we propose two open problems: show light
spanners for a planar graph with a single vortex, and show light
spanners for a path-like clique-sum of planar graphs.  For these cases
it may help to compose multiple charging schemes into an
``$\epsilon$-scheme'', as was necessary for apex
graphs~\cite{Grigni:2002:LSA:545381.545492}.  As usual, the main
difficulty is that we have no control over the MST topology; if the MST
has a nice topology (e.g.\ some form of monotonicity),
then we would be done.

Given a bounded treewidth graph, we can still define the notion of a
monotone spanning tree $T$.
We choose roots in the decomposition tree and $T$; whenever vertex $v$
has parent $p$ in $T$, we require $p$ to be in the bag containing $v$ which
is closest to the root.
If we can find such a $T$ that is light
enough, then we could repeat the rest of our argument from the bounded
pathwidth case.  However, there is an obstacle: the light monotone
tree might not exist.
\begin{theorem}\label{thm:nomono}
There is an edge-weighted graph $G$ with a bounded treewidth decomposition,
such that any monotone spanning tree $T$ in the completion of $G$
has weight $\Omega(\lg n)\cdot \MST(G)$.
\end{theorem}

\begin{figure}[ht]
\begin{center}
{\includegraphics[width=7cm]{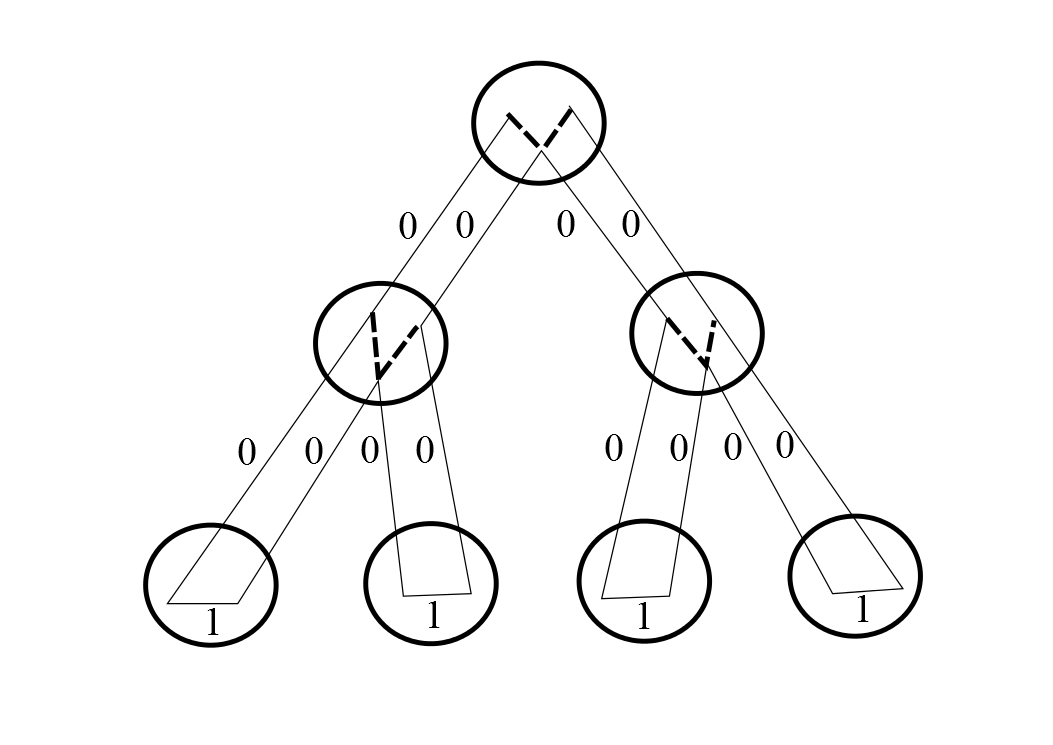}}
\caption{A bounded treewidth graph with no light monotone tree. 
The solid path $P$ is the minimum spanning tree. The horizontal edge
in each leaf bag has weight one, all other solid edges of $P$ have
weight zero.  All other edges (in particular, the dashed ones) have
weight equal to the distance in $P$ between its endpoints.}
\label{fig:treeex}
\end{center}
\end{figure}
\begin{proof}
We construct $G$ as follows (see Figure~\ref{fig:treeex}).   We start
with a balanced binary tree with $n$ nodes, think of this as our tree
of bags.  We assign 3 nodes to each internal bag, and 2 to each leaf bag.  
We connect these vertices by a path $P$ as shown; each edge of $P$ in a leaf
bag has weight one, all other edges in $P$ have weight zero.  (Note 
we must grow our bags a bit to support all these edges of $P$.)

Now in any monotone tree $T$ for $G$, for each internal bag, the
``bottom'' vertex of the three must be connected to one of the
other two (its parent in $T$); in other words, we must pick one of the
dashed edges shown in each internal bag.

The main observation is that if we sum up the weights of these
selected edges over one level of the decomposition tree, their total
is already a constant fraction of $w(P)$.  Summing over all levels,
the total weight $w(T)$ is $\Omega(\log n)\cdot w(P)$, as claimed.
\qed
\end{proof}


\bibliographystyle{plain}
\bibliography{ref}

\end{document}